\documentclass[12pt,a4paper]{article}
\usepackage[dvips]{graphics}
\usepackage{amsfonts,amsbsy}

\newcommand{\no}{\noindent}

\newcommand{\mod}[1]{\vert {#1}\vert}

\newcommand{\ket}[1]{\vert #1 \rangle}

\newcommand{\xb}{{\boldsymbol x}}
\newcommand{\yb}{{\boldsymbol y}}

\newcommand{\pb}{{\boldsymbol p}}

\newcommand{\intp}{\int\!\frac{d^3p}{(2\pi)^3}\;}

\newcommand{\bra}[1]{\langle #1 \vert}
\newcommand{\e}{{\mathrm{e}}}
\begin{document}


\begin{center}{\Large{\textbf{Effective Quarks and
Their Interactions}}}\footnote{Talk presented by D.~McMullan }\\ [8truemm]
\textsc{Emili Bagan\footnote{email: bagan@ifae.es}\\[5truemm]
\textit{
Physics Department\\
 Brookhaven National Laboratory\\
  Upton NY 11973\\
  USA}
\\[10truemm]
Robin Horan\footnote{emails: rhoran@plymouth.ac.uk,
mlavelle@plymouth.ac.uk and dmcmullan@plymouth.ac.uk }$\!\!$,
Martin Lavelle$^3$ and David McMullan$^3$}
\\
[5truemm] \textit{School of Mathematics and Statistics\\ The
University of Plymouth\\ Plymouth, PL4 8AA\\ UK} \end{center}

\bigskip\bigskip\bigskip
\begin{quote}
\textbf{Abstract:} This talk will summarise the progress we have made in our
programme to both characterise and construct charges in gauge
theories. As an application of these ideas we will see how the dominant
glue surrounding quarks, which is responsible for asymptotic
freedom, emerges from a constituent description of the interquark
potential.
\end{quote}

\bigskip\bigskip

\noindent\textbf{Introduction}

\bigskip
\noindent The phenomenological confidence in the existence of
coloured hadronic constituents is in marked contrast to the
theoretical uncertainties associated with attempts to describe
such charges. Although there are many  models of partonic and
effective degrees of freedom in the literature, none have yet
emerged directly from the fundamental gauge theoretic description
of the strong interactions
--- QCD.

The source of the difficulty in directly extracting these
effective degrees of freedom from the underlying gauge theory is a
particular example of the basic dichotomy we all face in QCD: the
degrees of freedom that make up the QCD Lagrangian and
successfully probe the ultra-violet regime are not related in any
obvious way to the large scale, infra-red degrees of freedom that
describe the observed hadrons or their constituents.

The picture that has emerged\footnote{For a nice discussion of the
stunning experimental results achieved over the past decade
see~\cite{Dainton:1999rq}.} from studies of deep inelastic scattering,
and more recently from diffractive processes, is that, as we probe
smaller and smaller sub-hadronic scales, we go from wholly
hadronic degrees of freedom to constituent structures (quarks). In
their turn, these constituents have properties (for example, their
mass) that run as shorter distance scales are probed. As such they
are not viewed as fundamental fields but as composites made up in
some way from the matter and gluonic degrees of freedom that enter
the QCD Lagrangian.

There is an  immediate theoretical problem for any coloured
constituent: how does such a quark or gluon have a well defined
colour given that, as is shown in the picture below, it is made up
of some mixture of apparently coloured partonic degrees of
freedom?

\begin{center}
\scalebox{0.5}{\includegraphics*{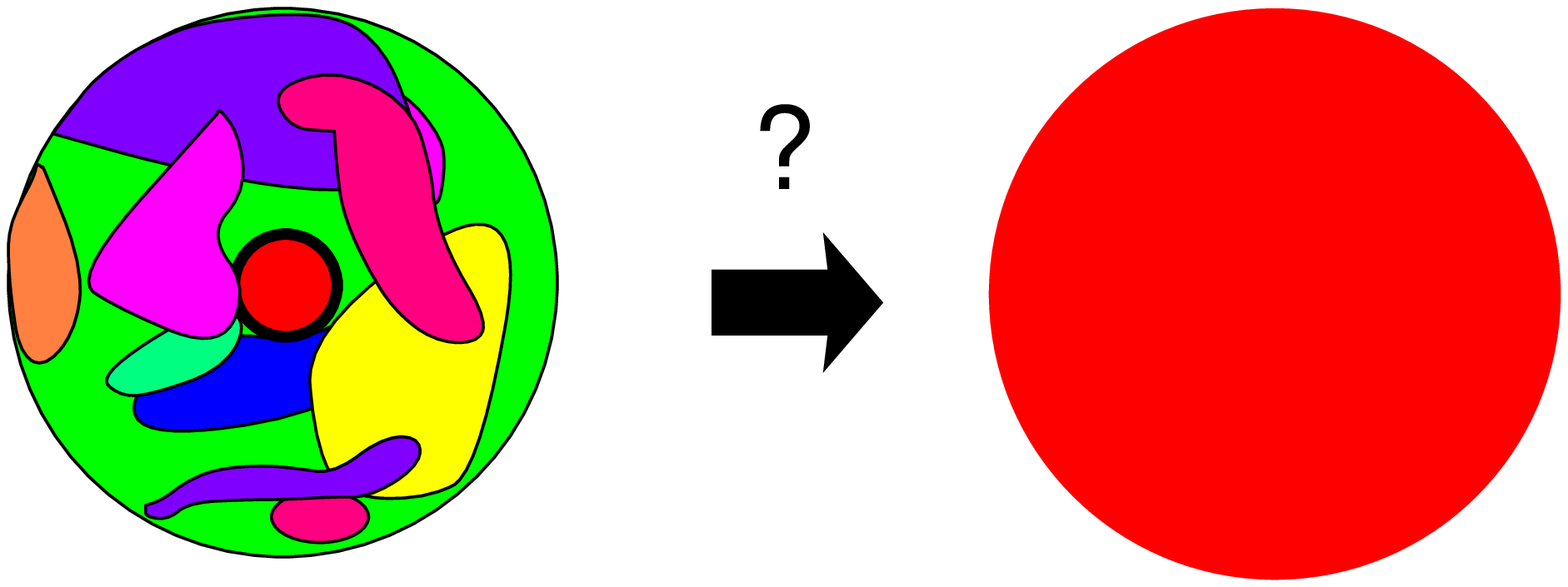}}
\end{center}
In order to answer this fundamental question and hence introduce
our approach~\cite{Lavelle:1997ty} to the construction of such
(colour) charged degrees of freedom, we need to recall \emph{how}
and \emph{when} colour can be defined in QCD.

\bigskip\bigskip

\goodbreak
\noindent\textbf{Colour and Gauge Invariance}
\bigskip

\noindent The very structure of QCD as a gauge theory tells us
that physical fields \emph{must} be invariant under
local\footnote{In QCD we actually expect the stronger statement to
emerge that physical fields are invariant under \emph{all} gauge
transformations:  both local and global, in other words colour
charges are supposed to be confined in (hadronic) colour singlets.
We will see at the end of this article how this stronger result
actually emerges directly from our approach to the constructions
of colour charges.} gauge transformations. This apparent truism
generates though an immediate problem when we look at the colour
charge itself:
\begin{equation}
Q^a=\int\! d^3x\,(J^a_0(x)-f^a_{bc}E^b_i(x)A^c_i(x))\,.
\end{equation}
This is clearly not gauge invariant! So can we talk about colour
in any meaningful way? The answer~\cite{Lavelle:1996tz} can be
seen to be yes, when we recognise that the question we should be
asking is whether the colour charge is gauge invariant when
restricted to physical (i.e., gauge invariant) states. On these
states, the non-abelian version of Gauss' law implies that
\begin{equation}
Q^a=\frac1{g}\int\! d^3x\,\partial_i E_i^a(x)\,.
\end{equation}
Under a gauge transformation $E_i^aT^a\to U^{-1}E_i^aT^aU$ so that
\begin{equation}
Q^aT^a\to\frac1{g}\int\! d^3x\,\partial_i (U^{-1}E_i^aT^aU)\,.
\end{equation}
We can now write this as the surface integral of the chromo-electric flux
in any given direction and hence we see that, on  gauge invariant
states, the colour charge transforms as
\begin{equation}
Q^aT^a\to
\frac1{g}\lim_{R\to\infty}\int_{S^2}d\underline{s}\cdot
U^{-1}\underline{\mathcal{E}}U\,.
\end{equation}
Hence the colour charge will be gauge invariant under local gauge
transformations  if, at spatial infinity, we have  $U\to U_\infty$
where
 $U_\infty$ lies in the centre of SU(3). Continuity then tells us
 that  this group element will be a constant independent of the
 direction taken to spatial infinity. This imposes a ${\mathbb Z}_3$
 (triality)  structure on possible charged states. However, we
 will only concern ourselves with the zero triality sector in this
 talk where $U_\infty$ is the identity.
To summarise the above discussion, we have seen that in order to
be able to define a coloured object, such as a quark, we require
that it  must be gauge invariant and also that allowed gauge
transformations must be restricted as above. We now want to study
the construction of charged fields.

\bigskip\bigskip

\goodbreak
\noindent\textbf{Construction of Charges}
\bigskip

\noindent In QCD we have the gauge transformations
\begin{eqnarray*}
A_\mu(x)&\to&  U^{-1}(x)A_\mu(x) U(x)+
\frac1g U^{-1}(x)\partial_\mu U(x)
\\
\psi(x)&\to& U^{-1}(x)\psi(x)
\end{eqnarray*}
From the non-triviality of these transformations we see that
neither of these fields have a well defined colour and hence they
\emph{cannot} be identified with observed gluonic or quark degrees
of freedom.

The generic form for a charged (matter) field is given by a
process we call \emph{dressing} the matter to give the product
\begin{equation}
h^{-1}(x)\psi(x)\,,
\end{equation}
where, under a gauge transformation, the dressing transforms as
\begin{equation}
h^{-1}(x)\to h^{-1}(x)U(x)\,.
\end{equation}
This is the minimal condition we must impose on the dressing in
order for the corresponding charged matter field to be gauge
invariant and hence having a well defined
colour~\cite{Lavelle:1997ty}. But does this really means we have a
physical field? Is gauge invariance alone enough? Consider the
stringy gauge invariant  $e^{+}e^{-}$ state
\begin{equation}
\ket{\bar\psi(x)\exp(ie\int_x^ydwA(w)) \psi(y)}
\end{equation}
This is gauge invariant, but is it physical? For static matter,
the  energy  is the expectation value of the Hamiltonian
$\frac12\int d^3z (E^2(z)+B^2(z))$. This yields for the  potential
energy of this state the confining potential
\begin{equation}
V(x-y)\sim e^2\mod{x-y}\delta^{2}(0)\,.
\end{equation}
Given that we are here dealing with four dimensional QED, it is
clear that the original stringy state cannot be accepted as a
physical configuration! Indeed it is an infinitely excited state
and what we would want to construct is the ground state for the
system. To do this in a systematic fashion, we need a further
condition apart from gauge invariance. The question which we now
address is: what condition on the dressing gives a stable charged
particle?

In order to motivate this extra condition, we first consider
${\varphi}$ to be a heavy field which creates a particle at the
point ${x}$ with a given 4-velocity ${u}$. The field must be
constant along the trajectory of a particle moving with this
velocity which leads to the equations of motion:
\begin{equation}
{ u^\mu \partial_\mu\varphi(x)=0\,.}
\end{equation}
If ${\varphi}$ is now a heavy gauged field then its equation of
motion becomes
\begin{equation}
{ u^\mu D_\mu\varphi(x)=0\,.}
\end{equation}
A physical heavy coloured field can only emerge from dressing this
field, ${\phi=h^{-1}\varphi}$. If this is truly  a heavy field,
then it should furthermore satisfy the equation
\begin{equation}
{ u^\mu \partial_\mu\phi(x)=0\,.}
\end{equation}
This means that the dressing must satisfy the \emph{dressing equation}
\begin{equation}
u^\mu\partial_\mu(h^{-1})=gh^{-1}u^\mu A_\mu\,.
\end{equation}
One can in fact show that this equation applies to any theory with
massive charges~\cite{Horan:1998im}.

\bigskip\bigskip

\noindent\textbf{Electric Charges}

\bigskip

\noindent It is instructive to first study the dressing process in
the simpler case of QED. Our two inputs into the construction are
then, as we have just seen, gauge invariance $h^{-1}\to h^{-1}
\mathrm{e}^{-ie\theta}$ and our kinematical requirement (the
dressing equation)
\begin{equation}
 u^\mu\partial_\mu(h^{-1})=-ieh^{-1}u^\mu A_\mu
\end{equation}
The great advantage of QED is that we can explicitly
solve~\cite{Bagan:1997kg} these equations to obtain
\begin{equation}
{ h^{-1}=\mathrm{e}^{-ieK}\mathrm{e}^{-ie\chi}}\,,
\end{equation}
where
\begin{eqnarray}
&&K(x)=-\int_{\Gamma}d\Gamma(\eta+v)^\mu\frac{\partial^\nu
F_{\nu\mu}}{{\mathcal G}\cdot\partial}\,,\\
&&\chi(x)=\frac{{\mathcal G}\cdot A}{{\mathcal G}\cdot\partial}
\,,
\end{eqnarray}
with $\eta=(1,\underline{0})$, $v=(0,\underline{v})$, ${\mathcal
G}^\mu=(\eta+v)^\mu(\eta-v)\cdot\partial-\partial^\mu$ and where
$\Gamma$ is the trajectory of the particle.

We thus see that the dressing has two structures: a gauge
dependent part, $\chi$, which makes the whole charge gauge
invariant, and is thus in some sense minimal, and a further gauge
invariant part, $K$ which is needed (together with the precise
form of $\chi$) to satisfy the dressing equation. These structures
are reflected in physical calculations: $\chi$ removes soft
divergences in QED calculations and, as we will see, in QCD
generates the anti-screening interaction responsible for
asymptotic freedom. $K$ removes the phase divergences in the
on-shell Green's functions of QED\footnote{The cancellation of the
various IR divergences~\cite{Bagan:1999jk} will be presented in
the talk by Martin Lavelle.}.

For greater insight into these structures, let us consider the
specific case of a static charge, $v=0$. This is:
\begin{equation}
{ h^{-1}\psi(x)=\mathrm{e}^{-ieK}\mathrm{e}^{-ie\chi}\psi(x)}
\end{equation}
with now
{
\begin{eqnarray}
&&K(x)=-\int_{-\infty}^{x^0}dt\frac{\partial^\nu
F_{\nu0}}{{\nabla^2}}\\
&&\chi(x)=\frac{\partial_i A_i}{\nabla^2}
\end{eqnarray}
}
and where
\begin{equation}{
\frac1{\nabla^2}f(t,\underline{x}):=-\frac1{4\pi}\int\!
d^3y\frac{f(t,\underline{y})}{|\underline{x}-\underline{y}|}}
\end{equation}
The non-locality of any description of a charge is manifest here.
The minimal part part of the electromagnetic cloud around a static
charge was first found by Dirac~\cite{Dirac:1955ca}. The
additional structure does not affect the electric field of the
charge and was therefore not picked up by Dirac's original
argument. We will now show that such charged (dressed) matter is
free at large times and that we can so  recover a particle
description~\cite{Bagan:1999jf}.

To see why this is important, we now recall  that Kulish and
Faddeev~\cite{kulish:1970} showed that, at large times the matter
field is not free, but rather becomes\footnote{For more details of
this method and a refinement of their work, see the talk by Robin
Horan}
\begin{equation}{
\psi(x)\rightarrow\!\intp \frac{ {D(p,t) }}{\sqrt{2E_p}}\left\{
b(\pb,s)u^s(\pb)e^{-ip\cdot x} +\cdots\right\}}\,,
\end{equation}
where ${D}$  is a \emph{distortion factor}. This implies that
there is no particle picture. Of course since the coupling does
not asymptotically vanish $\psi$ is not gauge invariant even at
large times and so we should not expect to relate it to a physical
particle! However, when we extract the annihilation operator for
our \emph{dressed} field we obtain
\begin{equation}{
b(q)\left\{
1+e\!\!\int_{\mathrm{soft}}\!\frac{d^3k}{(2\pi)^3}
\left(\frac{V\cdot a}{V\cdot k}-\frac{q\cdot a}{q\cdot
k}\right)\e^{-itk\cdot q/E_q}-{\mathrm{c.c.}}\right\}} +O(e^2)\,,
\end{equation}
with ${V^\mu=(\eta+v)^\mu(\eta-v)\cdot k-k^\mu}$.

There are two corrections now: the usual one~\cite{kulish:1970}
from the interactions of the  matter field and another one from
the dressing. It is easy to show~\cite{Bagan:1999jf} that at the
right point on the mass shell, ${q=m\gamma(\eta+v)}$, these
distortions cancel! We thus see that our dressed matter
asymptotically corresponds to free fields and we regain a particle
picture.

\bigskip\bigskip

\noindent\textbf{Colour Charges}

\bigskip

\no After this construction of abelian charges, we now want to
proceed to the non-abelian theory. We recall that the
\emph{minimal} static dressing in QED was: ${\exp(-ie\chi)}$, with
${\chi={\partial_i A_i}/{\nabla^2}}$. This vanishes in Coulomb
gauge and this observation lets us generalise this dressing to QCD
where it may be extended to an arbitrary order in ${g}$ (see the
Appendix of~\cite{Lavelle:1997ty}. Indeed we can also extend to
non-static charges, but in the application that follows we require
static quarks.

In QCD we write the dressing as a perturbative expansion
\begin{equation}{
\exp(-ie\chi)\Rightarrow \exp(g\chi^aT^a)\equiv h^{-1}}
\end{equation}
with ${g\chi^aT^a=(g\chi_1^a+g^2 \chi_2^a+g^3\chi_3^a+\cdots)T^a}$

The dressing gauge argument mentioned above implies
\begin{equation}{
\chi_1^a=\frac{\partial_j A_j^a}{\nabla^2}\,;\quad
\chi_2^a=f^{abc}\frac{\partial_j}{\nabla^2}\left(
\chi_1^bA_j^c+\frac12(\partial_j\chi_1^b)\chi^c_1
\right)}
\end{equation}
etc. This can be extended to all orders in the coupling. However,
we will return below to the question of non-perturbative
solutions.

\bigskip\bigskip

\noindent\textbf{The Interquark Potential}

\bigskip

\noindent We now want to study the interaction energy of the
ground state in the presence of a matter field and its antimatter
equivalent~\cite{Lavelle:1998dv}. To construct this we can either
dress a quark and an antiquark separately or dress a single meson
in which there are no gauge invariant constituents. We will now
dress the quark fields and study the potential between them:
whether or not this gives the correct interaction energy, i.e.,
the potential, is a test of the validity of a constituent picture.
We recall from our discussion of the \lq stringy\rq\ state that
this is a sensitive test.
%

Our procedure is as follows: as sketched above we first extend our
expression for the minimally dressed quark to higher orders in
perturbation theory. We then take such minimally dressed
quark/antiquark states, $\bar\psi(y)h(y)h^{-1}(y')\psi(y')\ket0$,
and sandwich the Hamiltonian,
\begin{equation}
{H=\frac12\int\!(E_i^aE_i^a+B_i^aB_i^a)d^3x}
\end{equation}
between them. Using the standard equal-time commutators
\begin{equation}{[E_i^a(x),A_j^b(y)]_{\mathrm et}=i\delta^{ab}
\delta(\xb-\yb)}\,,
\end{equation}
we can then calculate the potential.

The lowest order result, i.e., at order $g^2$, is just the Coulomb
potential:
\begin{equation}
  V^{g^2}(r)=
  -\frac{g^2 N C_F}{4\pi r}\,,
\end{equation}
where ${r}$ is the separation of the matter fields. This is of
course just QED with coloured icing.

What about  QCD with non-abelian ingredients? Well at order $g^4$
we  need to calculate the minimal static dressing to order $g^3$.
This can be done with the above mentioned efficient algorithm. A
relatively simple calculation then yields for the potential at
$g^4$
\begin{equation}\label{anti}
V^{g^4}(r)=-\frac{g^4}{(4\pi)^2}\frac{NC_FC_A}{2\pi r}{4}\log(\mu
r) \,.
\end{equation}
What does this tell us about our dressed state?

We recall that the QCD potential~\cite{Schroder:1998vy} may be
extracted from a Wilson loop as follows:
\begin{equation}{
V(r)=-\lim_{t\to\infty}
\frac1{it}\log\bra0\mathrm{Tr}\,\mbox{\sf{P}}\exp\left(
g\oint dx_\mu A^\mu_aT^a\right)\ket0}
\end{equation}
At order ${g^4}$ this yields
\begin{equation}{
V(r)= -\frac{g^2 C_F}{4\pi r}\left[ 1+
\frac{g^2}{4\pi}\frac{C_A}{2\pi }\left( {4}-\frac13 \right)
\log(\mu r) \right] }\,.
\end{equation}
From this we may read off the universal one-loop beta function
\begin{equation}{
\beta(g)=-\frac{g^3}{(4\pi)^2}
\left[{ 4}-{\frac13}\right]}\,.
\end{equation}
We have decomposed it in this way because it has been shown by a
number of
authors~\cite{Nielsen:1978rm,Hughes:1980ms,Hughes:1981nw,Drell:1981gu,Brown:1997nz}
that the dominant  anti-screening contribution (the 4) which is
responsible for asymptotic freedom comes from longitudinal glue
and the screening part (the $\frac13$) from gauge invariant glue

We now recognise that the result (\ref{anti}) for the interaction
energy of the minimally dressed quark/antiquark system is just the
anti-screening contribution to the interquark potential. We thus
make the important identification~\cite{Lavelle:1998dv} that the
\emph{dominant part of the glue in a $Q\bar Q$~system which is
responsible for anti-screening actually factorises into two
individually gauge invariant constituents}.  The success of
constituent models can, to the extent that these low order
calculations have been carried though, be explained by our work.
We postulate that the introduction of gauge invariant glue, via
the incorporation of the phase dressings, will produce the
screening effect.

\bigskip\bigskip

\noindent\textbf{Topology and Confinement}

\bigskip

\noindent Having seen the perturbative efficiency and relevance of
our variables, we now want to study the non-perturbative sector.
Experimentally of course we do not see free quarks, and it can be
easily shown that this is in fact predicted by our method.

This follows from an intimate link between dressings and
gauge-fixing. In the limited space available this can be best
explained pictorially. Essentially the existence of a gauge
fixing, $\chi$, can be shown~\cite{Lavelle:1997ty} to imply the
existence of a function $h$ which transforms as a minimal dressing
must to make a quark gauge invariant
\begin{center}
\scalebox{.5}{\includegraphics*{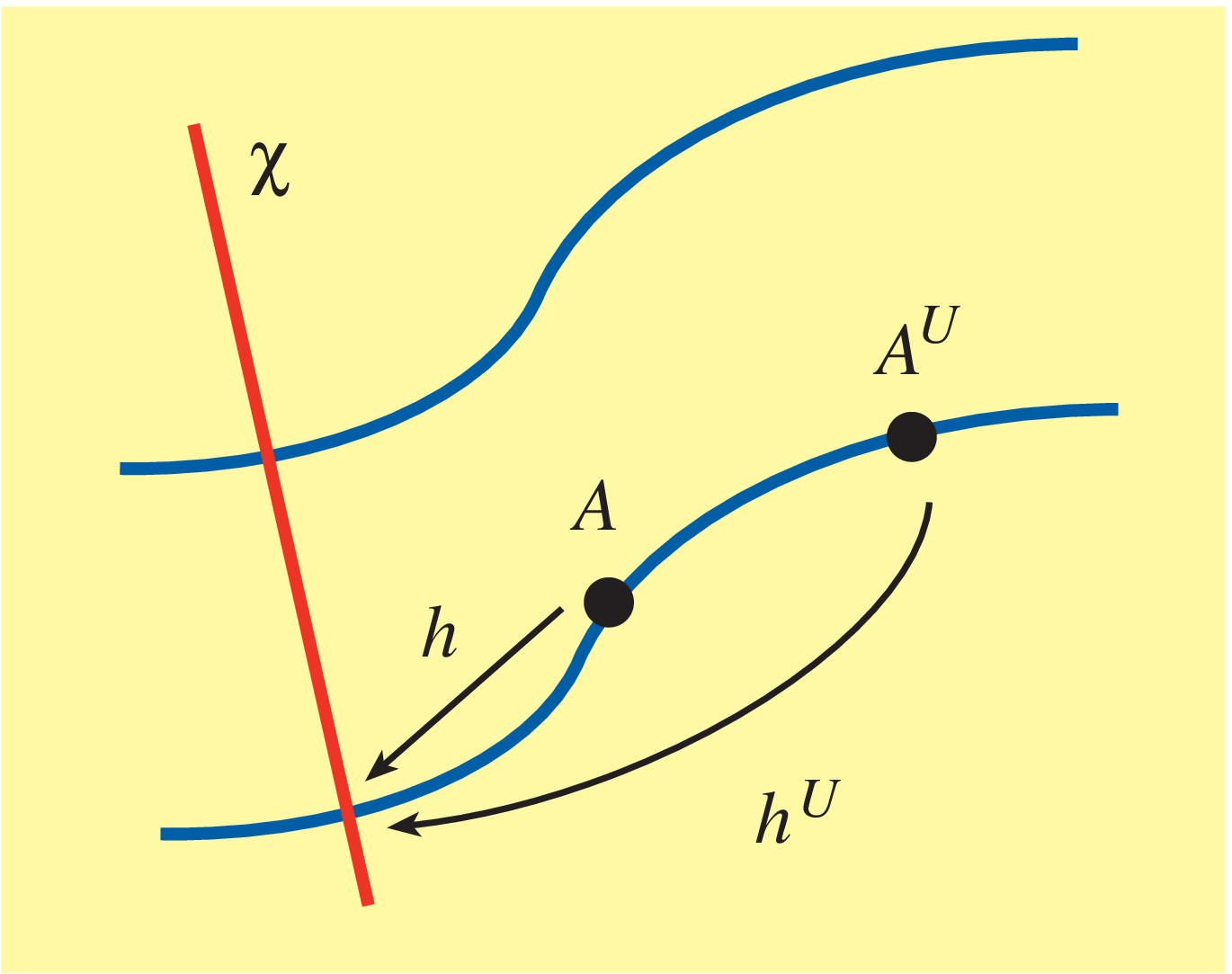}}
\end{center}
Similarly we can show~\cite{Lavelle:1997ty} that having a dressing
implies that we can construct a gauge fixing which slices every
gauge orbit once:
\begin{center}
\scalebox{.5}{\includegraphics*{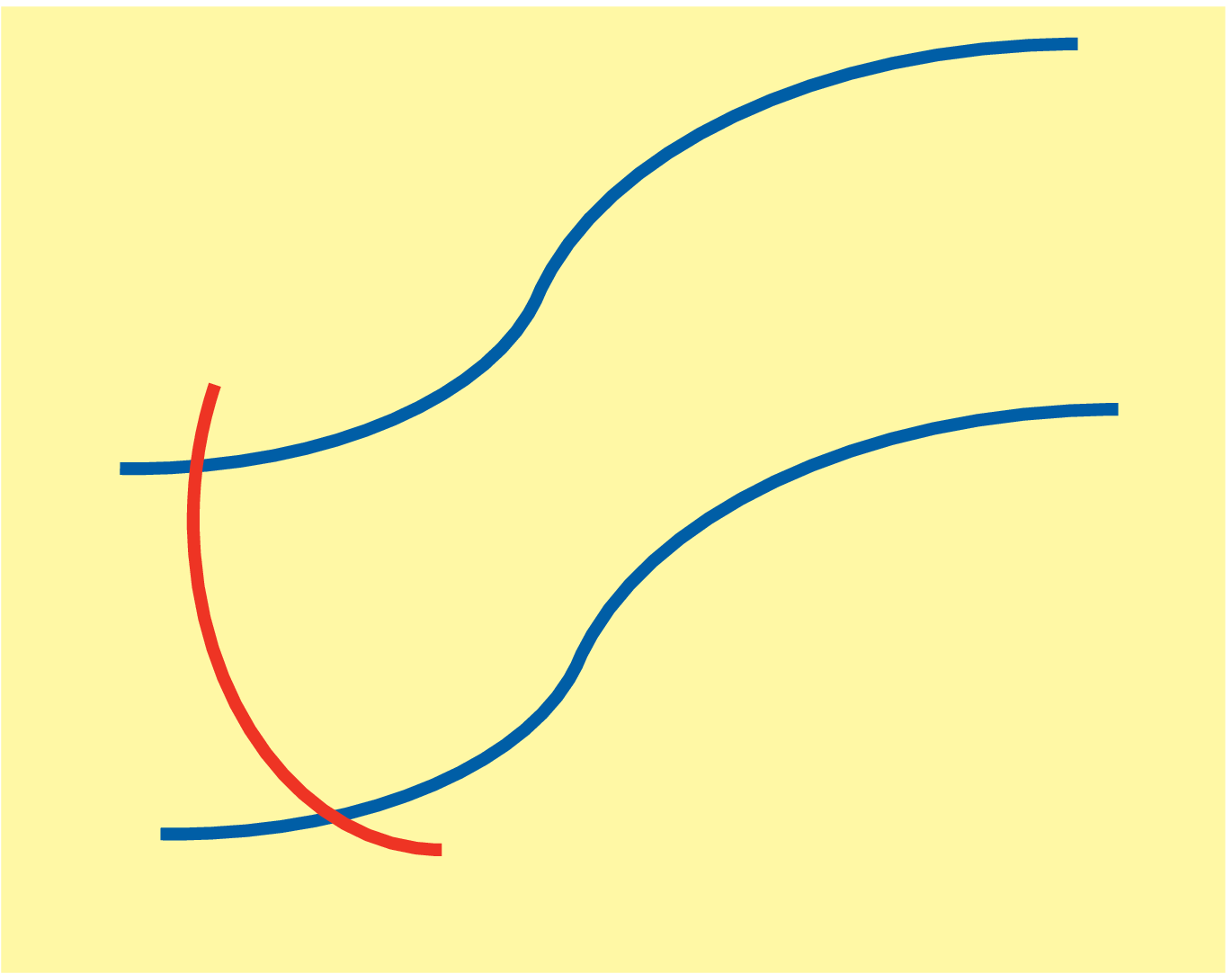}}
\end{center}
But Gribov and Singer~\cite{Gribov:1978wm,Singer:1978dk} have
showed that, with certain boundary conditions which we have seen
above are needed if colour is to be a good quantum number, there
is no such good gauge fixing in QCD!
\begin{center}
\scalebox{.5}{\includegraphics*{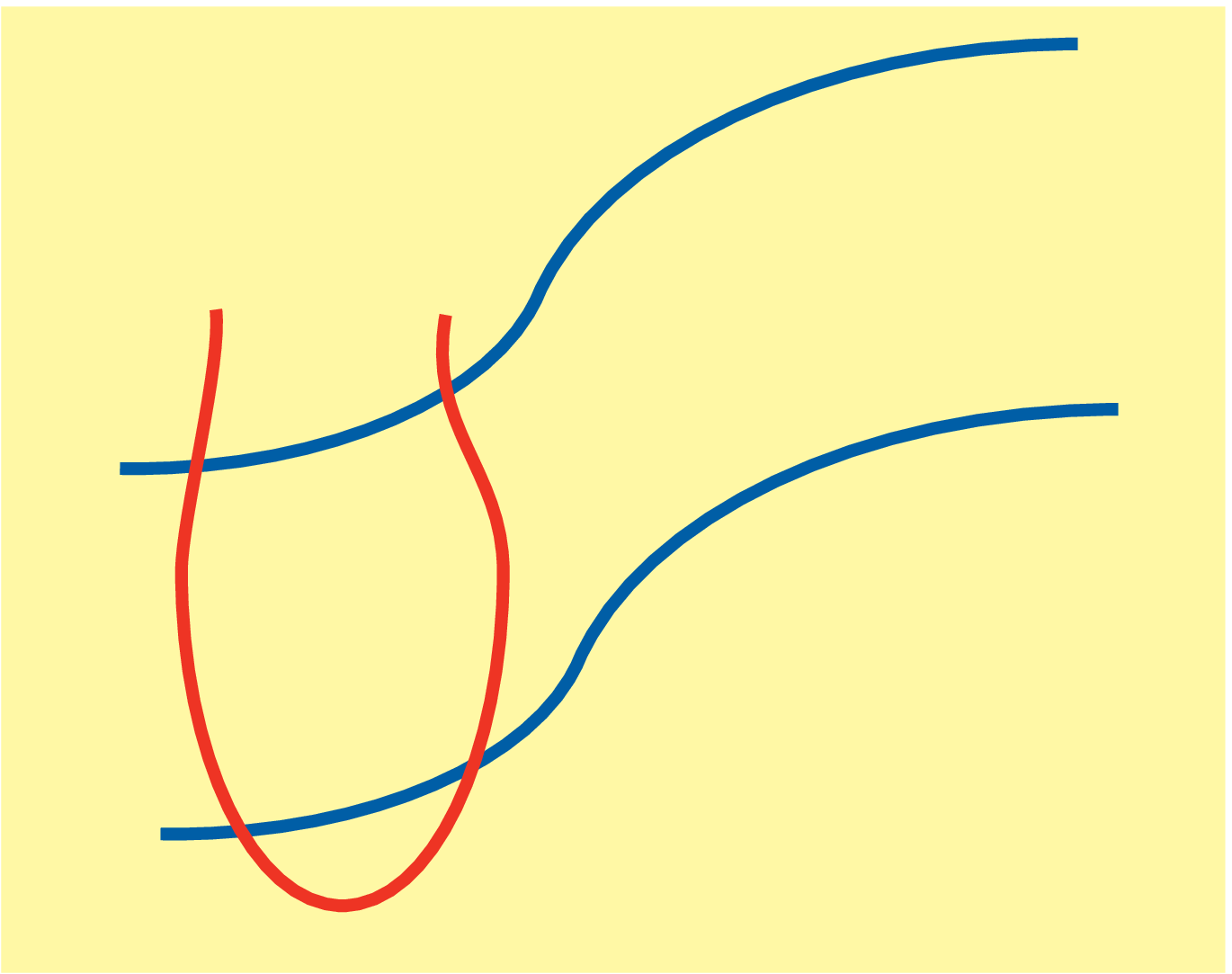}}
\end{center}
Hence any constituent picture must break down! We conclude that
\emph{there is no non-perturbative, gauge invariant description of
a single quark or gluon.} Therefore outside of certain dynamical
domains we cannot expect to see individual quarks or gluons.

\bigskip\bigskip

\noindent\textbf{Conclusions}

\bigskip

\noindent There are strong phenomenological reasons in both QED
and QCD for wanting to be able to describe charged particles. We
saw~\cite{Lavelle:1997ty} that if we want to describe coloured
substructure then this requires both gauge invariance and some
restrictions on the allowed gauge transformations.

We have seen a method of constructing dressed charges which was
designed to describe physical charges with a well defined
velocity. It had two inputs: gauge invariance and a kinematical
requirement to single out which of the potentially many gauge
invariant constructions involving a single matter field
corresponds to the ground state. This approach, we have shown,
generates structured dressings around charged particles.

It has long been argued that it is impossible to describe charged
particles in gauge theories~\cite{kulish:1970}. Essentially this
is because asymptotic matter fields are not free fields due to the
long range nature of the interactions transmitted by massless
gauge bosons. However, we have demonstrated~\cite{Bagan:1999jf}
that asymptotic dressed fields are indeed free fields (the
dressing takes such effects into account). We thus have obtained
for the very first time a particle description of charges. It will
be shown elsewhere in this meeting that this
removes~\cite{Bagan:1999jk} the IR divergences in QED at the level
of on-shell Green's functions.

We have calculated the potential between two minimally dressed
quark fields. The anti-screening contribution to the interquark
potential was shown to be generated by the minimal dressing. We
have thus determined the dominant glue configuration around
quarks. We have seen that in a \lq meson\rq\ -- at least to the
level at which we calculate -- two separate, gauge invariant,
coloured objects are visible.

Furthermore there is a topological obstruction to the construction
of coloured charges. This means that we can directly demonstrate
the non-observability of individual quarks or gluons: outside the
domain of perturbation theory and some non-perturbative effects,
the Gribov ambiguity will show itself and there will be no locally
gauge invariant description of such objects. This shows a new way
to calculate the scale of confinement: we need to find out at
which stage it becomes impossible to factorise the dressing of,
say, a $Q\bar Q$ state into two individual charges. Beyond this
breakdown of the factorisation quarks do not exist in QCD.

\bigskip
\no\textbf{Acknowledgements:} This research was partly supported
by the British Council/Spanish Education Ministry \textit{Acciones
Integradas} grant no.\ Integradas grant 1801 /HB1997-0141. It is a
pleasure to thank both the local organisers for their hospitality
and also PPARC for a travel grant.

\end{document}